\documentclass[journal, 10pt]{IEEEtran}
\usepackage{amstext,amsmath,graphicx,epsfig,amssymb}
\usepackage{amsfonts,algorithm,algorithmic,color,enumerate}
\usepackage{mathrsfs,xcolor,amsthm}
\usepackage{cancel}
\renewcommand{\baselinestretch}{2.0}

\newtheorem{definition}{Definition}[section]
\newtheorem{theorem}{Theorem}[section]
\newtheorem{corollary}{Corollary}[section]
\newtheorem{lemma}{Lemma}[section]
\numberwithin{equation}{section}
\newtheorem{fact}{Fact}
\onecolumn

\begin{document}

% Title.
% -------------------------------------------------------------------------------------------------------------------------------
\title{The Necessary And Sufficient Condition for Generalized Demixing}
% -------------------------------------------------------------------------------------------------------------------------------

% Single address.
% -------------------------------------------------------------------------------------------------------------------------------
\author{Chun-Yen Kuo, Gang-Xuan Lin, and Chun-Shien Lu
\thanks{The authors are with Institute of Information Science, Academia Sinica, Taipei, Taiwan.
E-mail: r00221004@ntu.edu.tw, spybeiman@gmail.com, and lcs@iis.sinica.edu.tw. Corresponding author: C.-S. Lu.}
%This work was supported by National Science Council, Taiwan, under grants NSC 102-2221-E-001-002-MY and NSC 102-2221-E-001-022-MY2.}
}
% -------------------------------------------------------------------------------------------------------------------------------

\maketitle

	\begin{abstract}
		Demixing is the problem of identifying multiple structured signals from a superimposed observation. This work analyzes a
		general framework, based on convex optimization, for solving demixing problems. We present a new solution to
		determine whether or not a specific convex optimization problem built for generalized demixing is
		successful. This solution will also bring about the possibility to estimate the probability of success by the approximate kinematic formula.
	\end{abstract}


%\IEEEpeerreviewmaketitle

\section{Introduction}
\IEEEPARstart{A}{ccording} to the theory of convex analysis,
	% convex cones take over the central role that subspaces perform in linear algebra. 	
	convex cones have been exploited to express the optimality conditions for a convex optimization problem [\ref{HUL93}].
	In particular, Amelunxen \textit{et al.} [\ref{ALMT13}] present the necessary and sufficient conditions for the problems of basis pursuit (BP) and demixing to be successful.
	
Let $x_0\in\mathbb{R}^n$ be an unknown $k$-sparse vector with $k$ nonzero entries in certain domain, let $A$ be an $m\times n$ random matrix whose entries are independent standard normal variables, and let $z = Ax_0\in\mathbb{R}^m$ be the measurement vector obtained via random transformation by $A$.
	In regard to the basis pursuit (BP) problem, which is defined as:
	\begin{equation}
		(\mbox{BP}) \text{ minimize } \Vert x\Vert_1 \text{ subject to } z=Ax,
		\label{BP}		
	\end{equation}	
a convex optimization method was proposed by Chen \textit{et al.} [\ref{CDS01}] to solve the sparse signal recovery problem in the context of compressive sensing [\ref{Donoho06}] when $m < n$.
	% to solve compressed sensing problem, motivated by work in geophysics [\ref{CM73}, \ref{SS86}].\\

To explore whether BP has a unique optimal solution, Amelunxen \textit{et al.} [\ref{ALMT13}] start from the concept of conic integral.
	\begin{definition}(descent cone). [\ref{ALMT13}]
		The \textit{descent cone} $\mathscr{D}(f,x)$ of a proper convex function $f:\mathbb{R}^n\rightarrow\bar{\mathbb{R}}$ at the point $x\in\mathbb{R}^n$ is the conical hull of the perturbations that do not increase $f$ near $x$.
	\begin{equation}
		\mathscr{D}(f,x):=\displaystyle\bigcup_{\tau>0}\lbrace y\in\mathbb{R}^n:f(x+\tau y)\leq f(x)\rbrace .
	\end{equation}
	\end{definition}
	%The descent cones of a proper convex function are always convex, but they may not be closed.
	%Also the descent cones of a smooth convex function are always halfspaces.
	%Thus, this concept inspires the most interest when the function is nonsmooth.

\begin{figure}[!hbpt]
\centering
			\vspace*{-10pt}\includegraphics[scale=0.4]{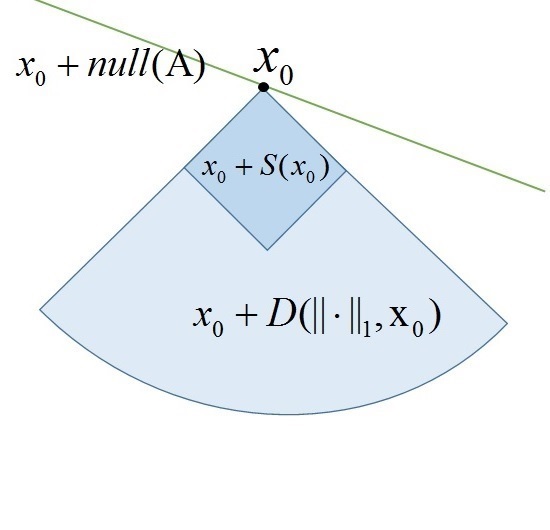}
			\hspace*{+15pt}
			\includegraphics[scale=0.4]{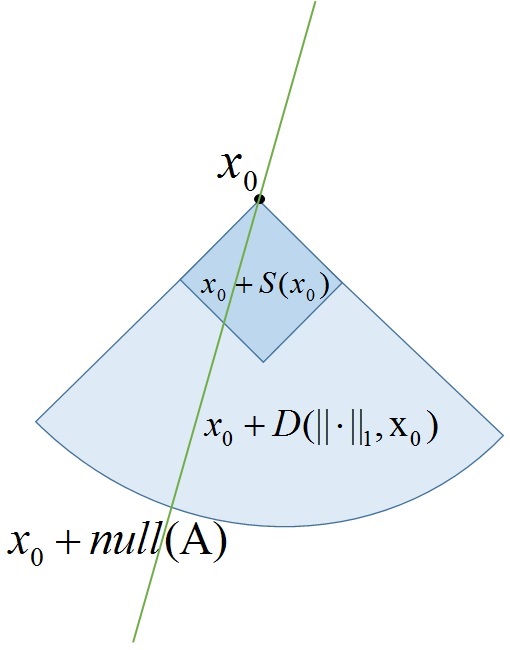}
		\caption{\textbf{The optimality condition for the BP problem.}
\textbf{[Left]}  BP succeeds. \textbf{[Right]} BP fails.
$S(x_0)=\lbrace y\in\mathbb{R}^d:\Vert x_0+y\Vert_1\leq\Vert x_0\Vert_1\rbrace$.
}
	\label{Fig: OC-BP}
	\end{figure}

We say that problem BP defined in Eq. (\ref{BP}) succeeds when it has a unique minimizer $\hat{x}$ that coincides with the true unknown, that is, $\hat{x}=x_0$.
	To characterize when the BP problem succeeds, Amelunxen \textit{et al.} present the primal optimality condition as:
	\begin{equation}\label{conBP}
		null(A)\cap\mathscr{D}(\left\|\cdot\right\|_1,x_0)=\lbrace 0\rbrace
	\end{equation}
	in terms of the descent cone [\ref{ALMT13}] (cf., [\ref{CRPW12}] and [\ref{RV08}]), where $null(A)$ denotes null space of $A$.
	The optimality condition for the BP problem is also illustrated in Fig. \ref{Fig: OC-BP}.

\begin{figure}[!hbpt]
\centering
			\includegraphics[scale=0.4]{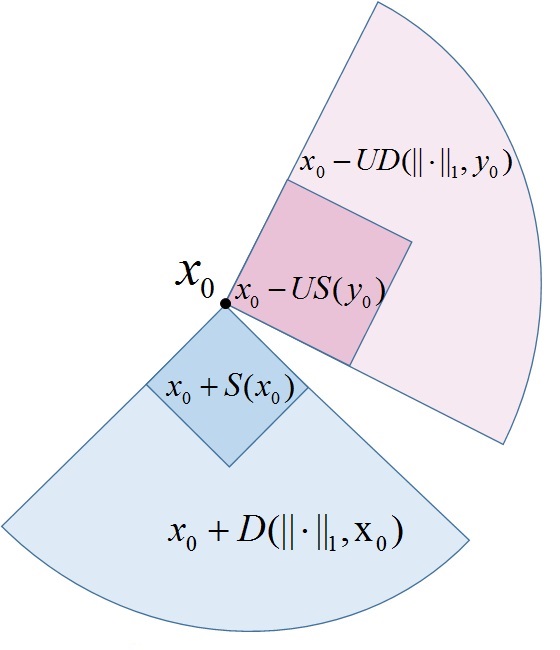}
\hspace*{+5pt}
		  \includegraphics[scale=0.4]{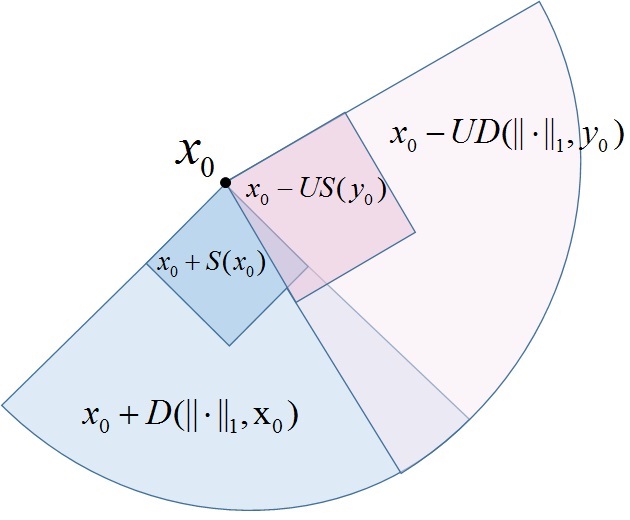}
		\caption{\textbf{The optimality condition for (DP) problem.}
		\textbf{[Left]} problem (DP) succeeds. \textbf{[Right]} problem (DP) fails.}
		\label{Fig: OC-Demixing}
	\end{figure}

Amelunxen \textit{et al.} [\ref{ALMT13}] also explore the demixing problem (sparse $+$ sparse) characterized as 
\begin{eqnarray}\label{demixing}
\begin{array}{c}
z=x_0+Uy_0,
\end{array}
\end{eqnarray}
where $U\in\mathbb{R}^{n\times n}$ is a known orthogonal matrix, and $x_0$ is itself sparse and $y_0$ is sparse with respect to $U$.
The optimization problem of recovering signals $x_0$ and $y_0$ is
formally defined as follow, which we call demixing problem (DP) in short:
	\begin{eqnarray}\label{s+s}
(\mbox{DP}) \left\{
	\begin{array}{l}
		\text{minimize } \Vert x\Vert_1 \\
		\text{subject to } \Vert y\Vert_1 \leq \Vert y_0\Vert_1 \text{ and } z=x+Uy.
	\end{array}\right.
	\end{eqnarray}
	They propose the primal optimality condition (also illustrated in Fig. \ref{Fig: OC-Demixing}) as:
	\begin{equation}\label{con s+s}
		\mathscr{D}(\left\|\cdot\right\|_1,x_0)\cap -U\mathscr{D}(\left\|\cdot\right\|_1,y_0)=\lbrace 0\rbrace
	\end{equation}
	to characterize whether $(x_0,y_0)$ is the unique minimizer to problem (DP).
	
%	\begin{eqnarray*}
%		\begin{array}{l}
%			null(A)\cap\mathscr{D}(\Vert\cdot\Vert_1,x_0) \\= null(A)\cap cone(S(x_0))\\ \ \\
%			\mathscr{D}(\Vert\cdot\Vert_1,x_0)\cap -U\mathscr{D}(\Vert\cdot\Vert_1,y_0) \\= cone(S(x_0))\cap -Ucone(S(y_0))
%		\end{array}
%	\end{eqnarray*}	
%	Besides, we can replace the $\mathscr{D}(\Vert\cdot\Vert_1,x_0)$ in the
%	condition (\ref{conBP}) with $S(x_0)$. Similarly, we can also replace the $\mathscr{D}(\Vert\cdot\Vert_1,x_0)$ and
%	$\mathscr{D}(\Vert\cdot\Vert_1,y_0)$ in the condition (\ref{con s+s}) with $S(x_0)$ and $S(y_0)$ respectively. That is,
%	\begin{eqnarray*}
%		\begin{array}{l}
%			null(A)\cap cone(S(x_0)) \equiv null(A)\cap S(x_0)\\ \ \\
%			cone(S(x_0))\cap -Ucone(S(y_0)) \equiv S(x_0)\cap -US(y_0)
%		\end{array}
%	\end{eqnarray*}		
%	\\
%	\
%	\\	
%	In other words, it is not necessary for them to exploit convex cone to express these primal optimality conditions.

The authors in [\ref{ALMT13}] also aim to estimate the probabilities of success of problem (BP) and problem (DP) with Gaussian random sensing matrices by the approximate kinematic formula.
They derive the probability\footnote{Nevertheless, the authors still fail to calculate the actual probabilities. In fact, they only derive the bounds of probabilities that involve the calculation of statistical dimension. Unfortunately, up to now the statistical dimension still cannot be calculated correctly.}
by using the convex (descent) cones.
Note that, as shown in Fig. \ref{Fig: OC-BP} and Fig. \ref{Fig: OC-Demixing}, the affine $\ell_1$ balls $S(\cdot)$ are defined as $S(a)=\lbrace x:\Vert x+a\Vert_1\leq\Vert a\Vert_1\rbrace$.
Also note that $cone(S(a))=\mathscr{D}(\left\|\cdot\right\|_1,a)$,
where $cone(S(a))$ is a conical hull of $S(a)$.

In Sections \ref{sec_motivation} and \ref{sec_main_result}, we generalize the demixing problem specified in Eq.
(\ref{demixing}),
set the corresponding optimization problem
to recover the signals of such generalized demixing model,
and explore its necessary and sufficient condition for successful demixing.

\section{Motivation and Problem Definition}\label{sec_motivation}

The demixing problem we discuss in this paper refers to the extraction of two informative signals from a single observation.
	We consider a more general model for a mixed observation $z\in\mathbb{R}^m$, which takes the form
	\begin{equation}
		z=Ax_0+By_0,
		\label{observation}
	\end{equation}
	where $x_0\in\mathbb{R}^{n_1}$ and $y_0\in\mathbb{R}^{n_2}$ are the unknown informative signals that we wish to find;
	the matrices $A\in\mathbb{R}^{m\times n_1}$ and $B\in\mathbb{R}^{m\times n_2}$ are arbitrary linear operators
	(not necessary $m\leq n_1$ or $n_2$).
	We assume that all elements appearing in Eq. (\ref{observation}) are known except for $x_0$ and $y_0$.
	The broad applications of the general model in Eq. (\ref{observation}) can be found in \cite{SB14} (and the references therein). 

It should be noted that: (1) if $y_0$ in Eq. (\ref{observation}) is set to zero, then the generalized demixing model is degenerated to BP; (2) The demixing model in
\cite{ALMT13} is a special case of Eq. (\ref{observation}) if $A$ is set to an identity matrix and $B$ is enforced to be an orthogonal matrix; (3) our generalized demixing model has more freedom
in the sense of dimension than that in \cite{MT13} because $A$ and $B$ can be arbitrarily selected.
Moreover, the two components $x_0$ and $y_0$ in our generalized model are permitted to have different lengths.

\section{Main Result}\label{sec_main_result}
The ground truths, $x_0$ and $y_0$, in Eq. (\ref{observation}) are approximated via solving the convex optimization problem defined as follows, which we call generalized demixing problem (GDP):
	\begin{eqnarray}\label{model}
	(\mbox{GDP})\left\{\begin{array}{l}
		\text{minimize } \Vert x\Vert_1 \\
		\text{subject to } \Vert y\Vert_1 \leq \Vert y_0\Vert_1 \text{ and } z=Ax+By.
	\end{array}\right.
	\end{eqnarray}
We call problem (GDP) succeeds provided $\left(x_0,y_0\right)$ is the unique optimal solution to GDP.
Our goal in this paper is to characterize when the problem (GDP) succeeds.

	\begin{theorem} \label{theorem1}
		The problem (\mbox{GDP}) has a unique minimizer $(\hat{x},\hat{y})$ to coincide with $(x_0,y_0)$
		\textbf{if and only if}
		\begin{eqnarray}\label{necessary1}
		\left\{
		\begin{array}{l}
		null(A)\cap S(x_0)=\lbrace 0\rbrace, \\
		null(B)\cap S(y_0)=\lbrace 0\rbrace, \\
		-AS(x_0)\cap BS(y_0)=\lbrace 0\rbrace.
		\end{array}\right.
		\end{eqnarray}
	\end{theorem}

\noindent \textbf{Proof:}
First, we assume that the problem (GDP) succeeds in having a unique minimizer $(\hat{x},\hat{y})$ to coincide with $(x_0,y_0)$.

\noindent
$\fbox{1}$ Claim: $null(A)\cap S(x_0)=\lbrace 0\rbrace$.\\
\indent Given $h_1 \in null(A)\cap S(x_0)$, we have $Ah_1 =0$.
By letting $(x',y')=(x_0 + h_1,y_0)$, it follows that
$z=Ax'+By'$ and $\left\|y'\right\|_1\leq\left\|y_0\right\|_1$,
which means that the point $(x',y')$ is a feasible point of problem (GDP).
On the other hand, since $h_1\in S(x_0)$, we have $\Vert x_0 + h_1\Vert_1=\left\|x'\right\|_1 \leq \Vert x_0 \Vert_1$.
By the fact that the problem (GDP) is assumed to have a unique minimizer $(x_0,y_0)$, we conclude that $h_1=0$.\\
\noindent
$\fbox{2}$ Claim: $null(B)\cap S(y_0)=\lbrace 0\rbrace$.\\
\indent Given $h_2 \in null(B)\cap S(y_0)$, we have $Bh_2 =0$ and $\Vert y_0 + h_2\Vert_1 \leq \Vert y_0 \Vert_1$.
By letting $(x'',y'') = (x_0,y_0+h_2)$, it follows that
$z=Ax''+By''$ and $\Vert y''\Vert_1 \leq \Vert y_0\Vert_1$.
Thus, $h_2=0$, otherwise $(x'',y'')\neq (x_0,y_0)$ will be another minimizer to problem (GDP).

\noindent
$\fbox{3}$ Claim: $-AS(x_0)\cap BS(y_0)=\lbrace 0\rbrace$.\\
\indent Given $s\in -AS(x_0)\cap BS(y_0)$, there exist $\bar{x}\in
S(x_0)$ and $\bar{y}\in S(y_0)$ to satisfy $-A\bar{x}=B\bar{y}=s$,
$\Vert x_0 + \bar{x}\Vert_1\leq\Vert x_0\Vert_1$, and
$\Vert y_0 + \bar{y}\Vert_1\leq\Vert y_0\Vert_1.$
By letting $(x''',y''')=(x_0+\bar{x},y_0+\bar{y})$,
it follows that $z=Ax'''+By'''$ and $\Vert y'''\Vert_1 \leq \Vert y_0\Vert_1$,
which mean that the point $(x''',y''')$ is a feasible point of problem (GDP).
On the other hand, since $\Vert x'''\Vert_1 \leq \Vert x_0\Vert_1$,
$(x''',y''')$ is also an optimal solution.
By the fact that the problem (GDP) is assumed to have a unique minimizer $(x_0,y_0)$,
%According to the assumption,
we conclude that $(x''',y''')=(x_0,y_0)$, and therefore $\bar{x}=0_{\mathbb{R}^{n_1}}$, $\bar{y}=0_{\mathbb{R}^{n_2}}$,
and $s=-A\bar{x}=0_{\mathbb{R}^{m}}$.

Conversely, we suppose the point $(x_0,y_0)$ satisfies Eq. (\ref{necessary1}).
Let $(x^*,y^*)$ be a feasible point of problem (GDP),
we aim to show that either $\Vert x^*\Vert_1 > \Vert x_0\Vert_1$ or
$(x^*,y^*)=(x_0,y_0)$.\\
\indent Let $h_1=x^*-x_0$ and $h_2=y^*-y_0$.
Since $(x^*,y^*)$ is feasible to problem (GDP), $z=Ax^*+By^*=A(x_0+h_1)+B(y_0+h_2)=z+Ah_1+Bh_2$, which implies $-Ah_1=Bh_2$.
If $\Vert x^*\Vert_1 > \Vert x_0\Vert_1$, then we are done.
So, we may assume $\Vert x^*\Vert_1 \leq \Vert x_0\Vert_1$, which means $h_1\in S(x_0)$.
Moreover, $\Vert y^*\Vert_1 \leq \Vert y_0\Vert_1$ implies
$h_2\in S(y_0)$.
Then, we get the fact that
$-Ah_1=Bh_2\in -AS(x_0)\cap BS(y_0)=\lbrace 0\rbrace$, namely,
$h_1\in null(A)$ and $h_2\in null(B)$.
Therefore, we have
$$h_1\in null(A)\cap S(x_0)=\lbrace 0\rbrace \mbox{ and } h_2\in null(B)\cap S(y_0)=\lbrace 0\rbrace,$$
which means $(x^*,y^*)=(x_0,y_0)$ and we complete the proof.
$\hfill\Box$
	
We know that $S(x_0)$ and $S(y_0)$ are the affine $\ell_1$-balls of the points $x_0$
and $y_0$, respectively. However, Eq. (\ref{necessary1}) is the formula, consisting of null spaces of
sensing matrices and affine $\ell_1$-balls.
Indeed we can relax the affine $\ell_1$-ball to be its conical hull such as
$\mathscr{D}(\left\|\cdot\right\|_1,x_0)=cone(S(x_0))$ and $\mathscr{D}(\left\|\cdot\right\|_1,y_0)=cone(S(y_0))$, and attain the following result.

\begin{corollary}
\label{theorem2}
		The problem (\mbox{GDP}) has a unique minimizer $(\hat{x},\hat{y})$ that coincides with $(x_0,y_0)$
		\textbf{if and only if}
		\begin{eqnarray}\label{necessary2}
		\left\{
		\begin{array}{l}
		null(A)\cap\mathscr{D}(\left\|\cdot\right\|_1,x_0)=\lbrace 0\rbrace, \\
		null(B)\cap\mathscr{D}(\left\|\cdot\right\|_1,y_0)=\lbrace 0\rbrace, \\
		-A\mathscr{D}(\left\|\cdot\right\|_1,x_0)\cap B\mathscr{D}(\left\|\cdot\right\|_1,y_0)=\lbrace 0\rbrace.
		\end{array}\right.
		\end{eqnarray}
\end{corollary}

We emphasize again that if $x_0$ and $y_0$ has the same length, as in the standard problem (Eq. \ref{demixing}), then $\mathscr{D}(\left\|\cdot\right\|_1,x_0)$ and $\mathscr{D}(\left\|\cdot\right\|_1,y_0)$ will reside in the same linear space and their intersection can be geometrically visible, as shown in Fig. \ref{Fig: OC-Demixing}.
However, since matrices $A$ and $B$ have arbitrary dimensions in our model, their geometrical interaction cannot simply be observed.
Thus, we argue that the derivation of necessary and sufficient condition via combining all of the cones is significantly different from standard problems \cite{ALMT13, MT13}.

\section{Simulations and Verifications}
We conduct simulations to verify the consistency between Theorem \ref{theorem1} and GDP.

\subsection{Verification Procedures}
The verification steps for practical sparse signal recovery based on Eq. (\ref{model}) are described as follows.

\begin{enumerate}
	\item[(1)] Construct the vectors $x_0\in\mathbb{R}^{n_1}$ and
$y_0\in\mathbb{R}^{n_2}$ with $k_1$ and $k_2$ nonzero entries, respectively.
The locations of the nonzero entries are selected at random, such nonzero entry equals $\pm 1$ with equal probability.
	\item[(2)] Draw two standard normal matrices $A\in\mathbb{R}^{m\times n_1}$ and $B\in\mathbb{R}^{m\times n_2}$, then capture the sample $z=Ax_0+By_0$.
 	\item[(3)] Solve problem (GDP) to obtain an optimal solution ($\hat{x}_1, \hat{y}_1$).
      \item[(4)] Declare successful demixing if $\Vert\hat{x}_1-x_0\Vert_2\leq 10^{-5 }$.
\end{enumerate}

\noindent
In addition, the verification steps for theoretic recovery based on Theorem \ref{theorem1} are described as follows.
\begin{enumerate}      
      \item[(5)] Solve $\min \Vert x_0+x\Vert_1$ subject to $Ax=0$ to obtain an optimal point $\hat{x}_2$.
      \item[(6)] Solve $\min \Vert y_0+y\Vert_1$ subject to $By=0$ to obtain an optimal point $\hat{y}_2$.
      \item[(7)] Solve $\min \Vert x_0+x\Vert_1$ subject to $Ax+By=0$ and $\Vert y_0+y\Vert_1\leq\Vert y_0\Vert_1$ to obtain a pair of optimal points $(\hat{x}_3,\hat{y}_3)$.
      \item[(8)] Declare success in Theorem \ref{theorem1} if $\ell_2$-norms of $\hat{x}_2$, $\hat{y}_2$, $\hat{x}_3$, and $\hat{y}_3$ are all smaller than or equal to $10^{-5 }$.
\end{enumerate}

\subsection{Simulation Setting and Results}
In our simulations, let $n_1$ and $n_2$ be the signal dimensions for signals $x_0$ and $y_0$, respectively.
Their sparsities, $k_1$ and $k_2$, ranged from $1$ to $n_1$ and $1$ to $n_2$, respectively.

\begin{figure}[!hbpt]
	\begin{center}
	\includegraphics[scale=0.09]{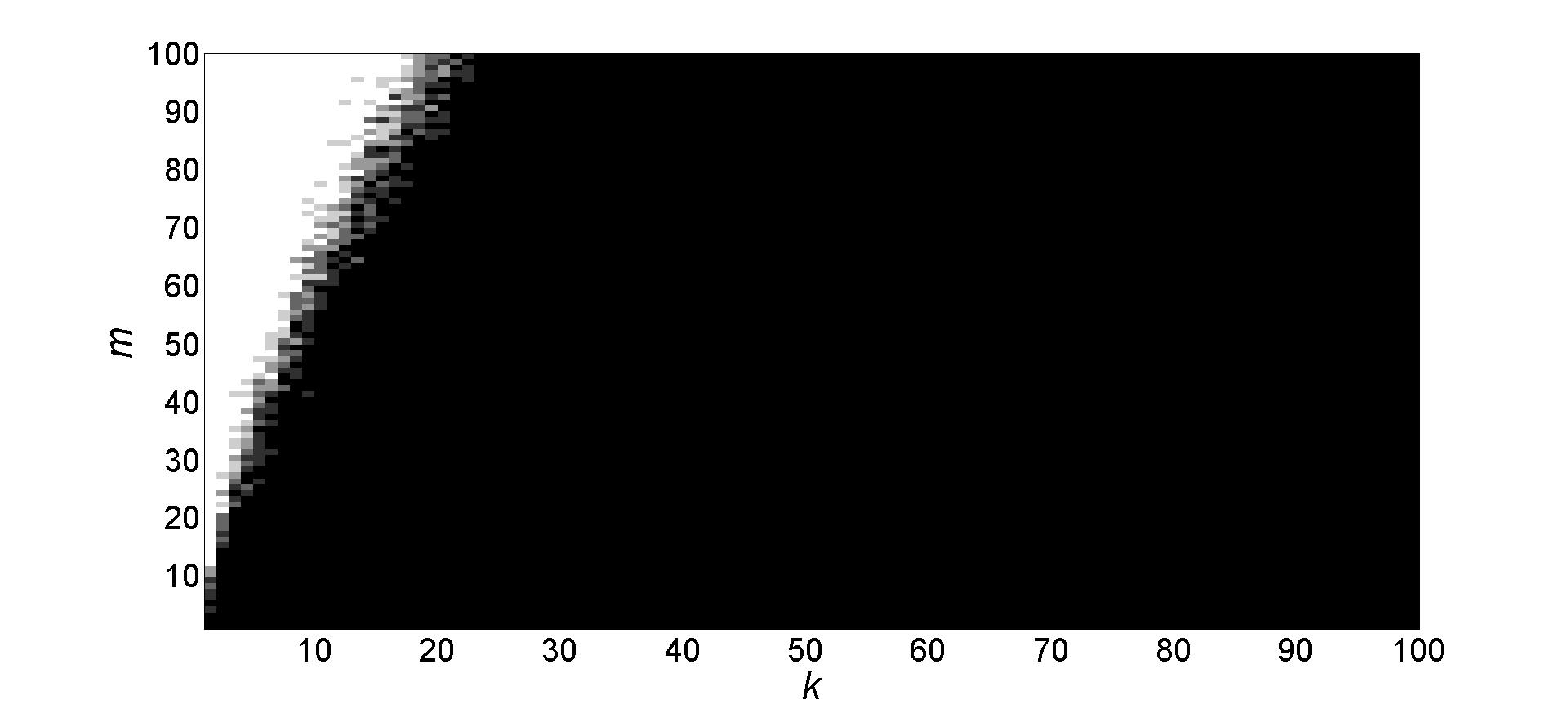}
	\includegraphics[scale=0.09]{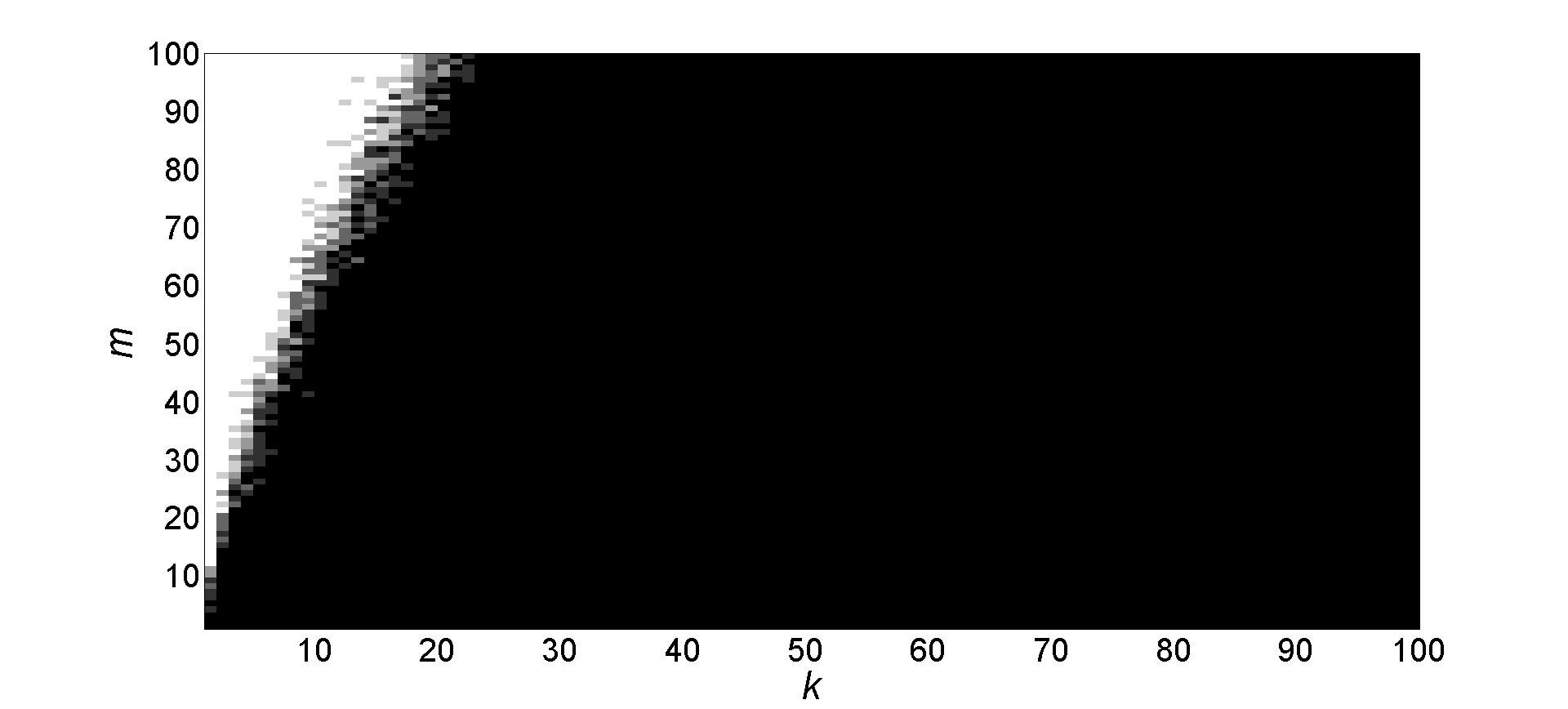}
	\end{center}
	\caption{
	\textbf{Phase transitions for demixing problems:
	[Left] Practical recovery of two sparse vectors based on Eq.
 (\ref{model}) and [Right] Theoretic recovery based on Theorem
 (\ref{theorem1}).}
	%Both show the probabilities that the demixing problem (\ref{model}) (Top) and Theorem (\ref{theorem1}) (Bottom) identify two sparse vectors $x_0, y_0\in\mathbb{R}^{100}$ given random linear measurements $z=Ax_0+By_0$.
	In each figure, the heat map indicates the empirical probability of success (black = $0\%$; white = $100\%$).
	}
	\label{Fig: Simulation Verification}
\end{figure}

First, we let $n_1=n_2=100$ and $k_1=k_2$.
Under the circumstance, the simulation results for both the demixing problems in Eq. (\ref{model}) and Theorem \ref{theorem1} are illustrated in Fig. \ref{Fig: Simulation Verification}, where the $x$-axis denotes the sparsity $k$ and the $y$-axis denotes the number $m$ of measurements.
We can see that the performances of these two seem to be identical and it is pretty easy to notice a fact that the smaller $k$ is, the easier for sparse signal recovery to succeed.

Second, we consider $n_1 \neq n_2$, where $n_1=100$ and $n_2=160$.
Again $k_1$ and $k_2$ ranged from $1$ to $n_1$ and $1$ to $n_2$, respectively.
By additionally considering varying number of measurements, the visualization of recovery results, unlike Fig. \ref{Fig: Simulation Verification}, will be multidimensional.
So, we chose different numbers of measurements with $10\leq m\leq 100$ in the simulations to ease observations.
The recovery result at each pair of $k_1$ and $k_2$ for each measurement rate $\left(\frac{m}{n}\right)$ was obtained by averaging from $100$ trials.
In sum, the simulation results reveal that, if each optimal solution in Steps (5)-(7) is zero, then the point, $x_0$ and $y_0$, satisfies Eq. (\ref{necessary1}), and vice versa.
That is to say, we can check if $x_0$ and $y_0$ satisfy Eq. (\ref{necessary1}) by solving these three optimization
problems in Steps (5)-(7).

\subsection{Proof of Feasibility of Our Verification}
Now we prove that why the above verification is feasible.
We say that $\hat{x}_2$, $\hat{y}_2$, $\hat{x}_3$, and $\hat{y}_3$ obtained from Steps (5)-(7) are all zero vectors if and only if  Eq. (\ref{necessary1}) in Theorem \ref{theorem1} holds.
We will validate this claim in the following.

\begin{definition}
Two cones $C$ and $D$ are said to \textbf{touch} if they share a ray but are weakly separable by a hyperplane.
\end{definition}
\begin{fact}\label{fact}
[\ref{SW08}, pp. 258-260]\\
Let $C$ and $D$ be closed and convex cones such that both $C$ and $D\neq \lbrace 0\rbrace$. Then
\begin{equation*}
	\mathbb{P}\lbrace \textbf{Q}C \text{ touches } D\rbrace =0,
\end{equation*}
where $\textbf{Q}$ is a random rotation.
\end{fact}

\begin{lemma}
Steps (5)-(8) constitute a complete verification to (\ref{necessary1}) in Theorem \ref{theorem1}.
\end{lemma}

\noindent
\textbf{Proof:} 
We want to prove that Steps (5)-(8) form a valid verification for
Eq. (\ref{necessary1}).
First, we assume the point $\left(x_0,y_0\right)$ satisfies
Eq. (\ref{necessary1}).\\
$\fbox{A1}$ Claim: $\hat{x}_2$ in Step (5) is zero.\\
\indent Since $\hat{x}_2$ is an optimal solution to the problem in Step (5),
we have $\left\|x_0+\hat{x}_2\right\|_1\leq\left\|x_0\right\|_1$ which implies $\hat{x}_2\in S(x_0)$; and
$A\hat{x}_2=0$ which is followed by $\hat{x}_2\in null(A)$.
That is $\hat{x}_2\in null(A)\cap S(x_0)=\left\{0\right\}$,
and hence $\hat{x}_2=0$.\\
$\fbox{A2}$ Claim: $\hat{y}_2$ in Step (6) is zero.\\
\indent The proof is similar to the one in \fbox{A1}.\\
$\fbox{A3}$ Claim: $\left(\hat{x}_3,\hat{y}_3\right)$ in Step (7) is zero.\\
\indent Since $\left(\hat{x}_3,\hat{y}_3\right)$ is an optimal solution to the problem
in Step (7), we have
$\left\|y_0+\hat{y}_3\right\|_1\leq\left\|y_0\right\|_1$ which means
$\hat{y}_3\in S(y_0)$;
$\left\|x_0+\hat{x}_3\right\|_1\leq\left\|x_0\right\|_1$
which says that $\hat{x}_3\in S(x_0)$; and
$A\hat{x}_3+B\hat{y}_3=0$ which implies
$-A\hat{x}_3=B\hat{y}_3\in-AS(x_0)\cap BS(y_0)=\left\{0\right\}$.
Thus, $\hat{x}_3\in null(A)$ and $\hat{y}_3\in null(B)$, then
$\hat{x}_3\in null(A)\cap S(x_0)=\left\{0\right\}$, $\hat{y}_3\in null(B)\cap S(y_0)=\left\{0\right\}$ and come to the conclusion
that $\left(\hat{x}_3,\hat{y}_3\right)=\left(0,0\right)$.

On the other hand, suppose that the optimal solutions $\hat{x}_2, \hat{y}_2,$
and $\left(\hat{x}_3,\hat{y}_3\right)$ corresponding to minimization problems in Steps (5), (6), and (7), respectively, are all zeros.

\begin{figure}[!hbpt]
\centering
\includegraphics[width = 0.30\textwidth]{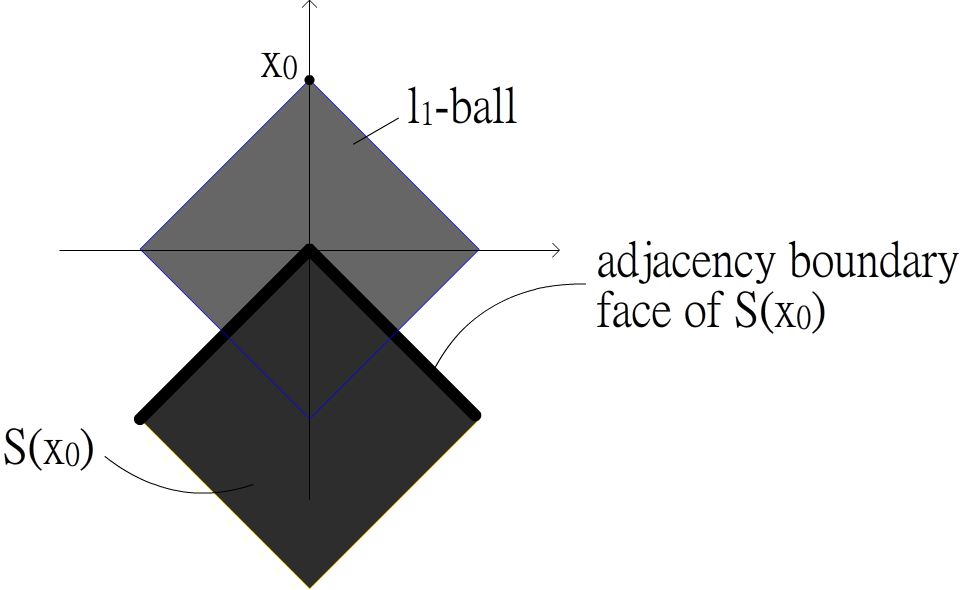}
\caption{The adjacency boundary face of $S(x_0)$.}\label{abf}
\end{figure}

\noindent
$\fbox{B1}$ Claim: $null(A)\cap S(x_0)=\lbrace 0\rbrace$.\\
\indent Given $x^*\in null(A)\cap S(x_0)$, we have $x^*\in S(x_0)$, meaning that 
% = \left\{x:\left\|x_0+x\right\|_1\leq\left\|x_0\right\|_1\right\}$,
$\left\|x_0+x^*\right\|_1\leq\left\|x_0\right\|_1$.
Furthermore, we also have $x^*\in null(A)$, which implies that $x^*$ is a feasible point of problem in Step (5).
Due to the fact that
$$\left\|x_0\right\|_1=\left\|x_0+\hat{x}_2\right\|_1\leq
\left\|x_0+x\right\|_1\ \forall x\in null(A),$$
we have $\left\|x_0+x^*\right\|_1\geq
\left\|x_0\right\|_1$. Thus $\left\|x_0+x^*\right\|_1=\left\|x_0\right\|_1$,
which means $x^*$  belongs to adjacency boundary face
$\partial_*(S(x_0))$ of $S(x_0)$ at $x_0$,
where adjacency boundary face
$\partial_*(\cdot)=\partial(\cdot)\cap\partial(cone(\cdot))$
is the intersection of boundary of itself and boundary of its conical hull
(as shown in Fig. \ref{abf}).
Therefore,
$null(A)$ touches $\mathscr{D}(\left\|\cdot\right\|_1,x_0)$ or $null(A)\cap S(x_0)=\lbrace 0\rbrace$.\\

By Fact \ref{fact}, we may assume that ``$null(A)$ touches $\mathscr{D}(\Vert\cdot\Vert_1,x_0)$'' never happens.
So we conclude that ``$null(A)\cap S(x_0)=\lbrace 0\rbrace$''.

\noindent
$\fbox{B2}$ Claim: $null(B)\cap S(y_0)=\lbrace 0\rbrace$.\\
\indent The proof is similar to the one in \fbox{B1}.

\noindent
$\fbox{B3}$ Claim: $-AS(x_0)\cap BS(y_0)=\left\{0\right\}$.\\
\indent Given $s^*\in -AS(x_0)\cap BS(y_0)$, there exist $x^*\in S(x_0)$ and $y^*\in S(y_0)$ such that $s^*=-Ax^*=By^*$.
Since $y^*\in S(y_0)$, we have
$\left\|y_0+y^*\right\|_1\leq\left\|y_0\right\|_1$, together with the fact that
$Ax^*+By^*=0$, the point $\left(x^*,y^*\right)$ is a feasible point of the problem is Step (7).\\
\indent Since $\left(\hat{x}_3,\hat{y}_3\right)=(0,0)$
is an optimal solution to problem in Step (7), we have
$$\left\|x_0\right\|_1=\left\|x_0+\hat{x}_3\right\|_1
\leq\left\|x_0+x^*\right\|_1.$$
Moreover, $x^*\in S(x_0)$ means
$\left\|x_0+x^*\right\|_1\leq\left\|x_0\right\|_1$.
Thus, $\left\|x_0+x^*\right\|_1=\left\|x_0\right\|_1$,
{\em i.e.}, $x^*\in\partial_*\left(S(x_0)\right)$ and $s^*\in\partial_*\left(-AS(x_0)\right)$.
Therefore, $$-AS(x_0)\cap BS(y_0)\subseteq \partial_*(-AS(x_0)),$$
which means
$-A\mathscr{D}(\left\|\cdot\right\|_1,x_0)$ touches $B\mathscr{D}(\left\|\cdot\right\|_1,y_0)$ or $-AS(x_0)\cap BS(y_0)=\lbrace 0\rbrace$.
Due to Fact \ref{fact}, we may assume that
``$-A\mathscr{D}(\left\|\cdot\right\|_1,x_0)$ touches $B\mathscr{D}(\left\|\cdot\right\|_1,y_0)$" never happens. So we conclude that
``$-AS(x_0)\cap BS(y_0)=\left\{0\right\}$".
$\hfill\Box$

\section{Future work}
We plan to employ Corollary \ref{theorem2} to estimate the probability of success under some assumptions by the approximate kinematic formula from [\ref{ALMT13}].
	\begin{theorem} \label{theoremAKF}
		(Approximate kinematic formula)\\
		Fix a tolerance $\eta\in (0,1)$. Let $C$ and $K$ be convex cones in $\mathbb{R}^n$, and draw a random orthogonal basis
		$Q\in\mathbb{R}^{d\times d}$. Then
		\begin{eqnarray*}
			\begin{array}{l}
				\delta (C)+\delta (K)\leq d-a_\eta\sqrt{d} \Rightarrow
				\displaystyle\mathbb{P}\lbrace C\cap QK\neq\lbrace 0\rbrace\rbrace\leq\eta\\
				
				\delta (C)+\delta (K)\geq d+a_\eta\sqrt{d} \Rightarrow
				\displaystyle\mathbb{P}\lbrace C\cap QK\neq\lbrace 0\rbrace\rbrace\geq 1-\eta,
			\end{array}
		\end{eqnarray*}
	  where $a_\eta :=\sqrt{8log(4/\eta)}$ and $\delta$ means the statistical dimension.
	\end{theorem}
	
	\begin{definition}(Statistical dimension)\\
		Let $C\subseteq\mathbb{R}^d$ be a closed convex cone.
		Define the Euclidean projection
		$\displaystyle\Pi_C:\mathbb{R}^d\rightarrow C$ onto $C$ by
		\begin{eqnarray*}
			\displaystyle\Pi_C(x):=\mathop{\arg\min}_{y\in C}\Vert y-x\Vert^2.
		\end{eqnarray*}	
		The statistical dimension $\delta (C)$ of $C$ is defined as:
		\begin{eqnarray*}
			\displaystyle\delta (C):=\mathbb{E}_g[\Vert\Pi_C(g)\Vert^2],
		\end{eqnarray*}		
		where $g\sim N(0,I)$ is a standard Gaussian vector.
	\end{definition}

	For the generalized demixing model proposed in this paper, we suppose $A\in\mathbb{R}^{m\times n_1}$ and $B\in\mathbb{R}^{m\times n_2}$ have independent standard normal entries, and let $z=Ax_0+By_0$.
	For the compressive sensing demixing, we may assume $m<n_1$, $m<n_2$, and both $A$ and $B$ have full rank.
	Then. we can derive:
		\begin{eqnarray}\label{Eq: Cone-Success}
			\left\{
			\begin{array}{l}
				m\geq \delta(\mathscr{D}(\Vert\cdot\Vert_1,x_0)) + a_{\eta_1}\sqrt{n_1}, \\
				m\geq \delta(\mathscr{D}(\Vert\cdot\Vert_1,y_0)) + a_{\eta_2}\sqrt{n_2}, \\
				m\geq \delta(A\mathscr{D}(\Vert\cdot\Vert_1,x_0))+\delta(B\mathscr{D}(\Vert\cdot\Vert_1,y_0))
				+ a_{\eta_3}\sqrt{m},
			\end{array}\right.
		\end{eqnarray}	
		which implies
		\begin{eqnarray*}
			\left\{
			\begin{array}{l}
				\mathbb{P}\lbrace null(A)\cap\mathscr{D}(\Vert\cdot\Vert_1,x_0)=\lbrace 0\rbrace\rbrace\geq 1-\eta_1, \\
				\mathbb{P}\lbrace null(B)\cap\mathscr{D}(\Vert\cdot\Vert_1,y_0)=\lbrace 0\rbrace\rbrace\geq 1-\eta_2, \\
				\mathbb{P}\lbrace -A\mathscr{D}(\Vert\cdot\Vert_1,x_0)\cap B\mathscr{D}(\Vert\cdot\Vert_1,y_0)=\lbrace 0\rbrace\rbrace\geq 1-\eta_3.
			\end{array}\right.
		\end{eqnarray*}

  \noindent
	On the other hand, we also have
		\begin{eqnarray}\label{Eq: Cone-Failure}
			\left\{
			\begin{array}{l}
				m\leq \delta(\mathscr{D}(\Vert\cdot\Vert_1,x_0)) - a_{\eta_1}\sqrt{n_1}, \\
				m\leq \delta(\mathscr{D}(\Vert\cdot\Vert_1,y_0)) - a_{\eta_2}\sqrt{n_2}, \\
				m\leq \delta(A\mathscr{D}(\Vert\cdot\Vert_1,x_0))+\delta(B\mathscr{D}(\Vert\cdot\Vert_1,y_0))
				- a_{\eta_3}\sqrt{m},
			\end{array}\right.
		\end{eqnarray}	
		which implies
		\begin{eqnarray*}
			\left\{
			\begin{array}{l}
				\mathbb{P}\lbrace null(A)\cap\mathscr{D}(\Vert\cdot\Vert_1,x_0)=\lbrace 0\rbrace\rbrace\leq \eta_1, \\
				\mathbb{P}\lbrace null(B)\cap\mathscr{D}(\Vert\cdot\Vert_1,y_0)=\lbrace 0\rbrace\rbrace\leq \eta_2, \\
				\mathbb{P}\lbrace -A\mathscr{D}(\Vert\cdot\Vert_1,x_0)\cap B\mathscr{D}(\Vert\cdot\Vert_1,y_0)=\lbrace 0\rbrace\rbrace\leq \eta_3.
			\end{array}\right.
		\end{eqnarray*}

Apparently, if the number $m$ of measurements is large enough, then successful sparse recovery can be achieved.
On the other hand, failed recovery is possible due to insufficient number of measurements.
But if we want to realize the above derived results, computation of the statistical dimensions of $A\mathscr{D}(\Vert\cdot\Vert_1,x_0)$ and $B\mathscr{D}(\Vert\cdot\Vert_1,y_0)$, as indicated in Eqs. (\ref{Eq: Cone-Success}) and (\ref{Eq: Cone-Failure}), will be an unavoidable difficulty.

\section{Conclusion}
Our major contribution in this paper is to derive the necessary and sufficient condition for a successful generalized demixing problem.
There is an issue worth mentioning, {\em i.e.}, Amelunxen \textit{et al.} have evaluated an upper bound and a lower bound of the probability of successful recovery for demixing problem (DP).
The reason why we did not do that is due to the known unavoidable difficulty raised by the generalized model (GDP problem), that is, \textbf{``How to compute the statistical dimension of a descent cone operated by a linear operator?"}.
We believe that if this open problem can be solved, we will complete the
generalized demixing problem with Gaussian random measurements.

\section{Acknowledgment}
This work was supported by National Science Council, Taiwan, under grants NSC 102-2221-E-001-002-MY and NSC 102-2221-E-001-022-MY2.

% -------------------------------------------------------------------------------------------------------------------------------
%\bibliographystyle{IEEEbib}
%%\bibliographystyle{abbrv}
%\baselineskip=9pt plus1pt
%\bibliography{Reference}

\begin{thebibliography}{1}

\bibitem{ALMT13}\label{ALMT13}
	D. Amelunxen, M. Lotz, M. B. McCoy, and J. A. Tropp.
	Living on the edge: A geometric theory of phase transitions in convex optimization.
	arXiv preprint arXiv:1303.6672, 2013.

\bibitem{MT13}\label{MT13}
	M. B. McCoy and J. A. Tropp.
	The achievable performance of convex demixing.
	arXiv preprint arXiv:1309.7478, 2013.
		
\bibitem{CRPW12}\label{CRPW12}
	V. Chandrasekaran, B. Recht, P. A. Parrilo, and A. S. Willsky.
	The convex geometry of linear inverse problems.
	Found. 	Comput. Math., 12(6):805-849, 2012.	
		
\bibitem{CDS01}\label{CDS01}
	S. S. Chen, D. L. Donoho, and M. A. Saunders.
	Atomic decomposition by basis pursuit. SIAM Rev., 43(1):129-159, 2001.
	Reprinted from SIAM J. Sci. Comput. 20 (1998), no. 1, 33-61 (electronic).
		
\bibitem{CM73}\label{CM73}
	J. F. Claerbout and F. Muir.
	Robust modeling of erratic data.
	Geophysics, 38(5):826-844, October 1973.

\bibitem{Donoho06}\label{Donoho06}
  D. L. Donoho.
  Compressed sensing
  IEEE Trans. on Information Theory, 52(4): 1289-1306, 2006.
	
\bibitem{HUL93}\label{HUL93}
	J.-B. Hiriart-Urruty and C. Lemarchal.
	Convex analysis and minimization algorithms.
	I, volume 305 of Grundlehren der Mathematischen Wissenschaften [Fundamental Principles of Mathematical Sciences].
	Springer-Verlag, Berlin, 1993. Fundamentals.

\bibitem{RV08}\label{RV08}
	M. Rudelson and R. Vershynin.
	On sparse reconstruction from Fourier and Gaussian measurements.
	Comm. Pure Appl. Math.,61(8):1025-1045, 2008.
	
\bibitem{SB14}\label{SB14}
	C. Studer and R. G. Baraniuk.
	Stable restoration and separation of approximately sparse signals.
	Applied and Computational Harmonic Analysis, 37:12-35, 2014.
	
\bibitem{SS86}\label{SS86}
	F. Santosa and W. W. Symes.
	Linear inversion of band-limited reflection seismograms.
	SIAM J. Sci. Statist. Comput.,7(4):1307-1330, 1986.

\bibitem{SW08}\label{SW08}
	R. Schneider and W. Weil.
	Stochastic and Integral Geometry.
	Springer series in statistics: Probability and its applications. Springer, 2008.
	
\end{thebibliography}

\end{document}